\documentclass[aip,jcp,reprint,groupedaddress]{revtex4-1}
\usepackage{graphicx,amssymb,amsmath,comment}

\begin{document}

%\preprint{}


\title{Electro-Osmotic Flow of Semidilute Polyelectrolyte Solutions}


\author{Yuki Uematsu} 
\email{y\_uematsu@scphys.kyoto-u.ac.jp}
\author{Takeaki Araki}
\affiliation{Department of Physics, Kyoto University, Kyoto 606-8502, Japan}


\date{\today}

\begin{abstract}
We investigate electro-osmosis in 
aqueous solutions of polyelectrolytes 
using mean-field equations. 
A solution of positively charged polyelectrolytes 
is confined between two negatively charged planar surfaces, 
and an electric field is applied parallel to the surfaces. 
When electrostatic attraction between the polymer and the surface is strong, 
the polymers adhere to the surface, forming a highly viscous adsorption layer that greatly suppresses the electro-osmosis. 
Conversely, electro-osmosis is enhanced by depleting the polymers from the surfaces. 
We also found that the electro-osmotic flow is invertible
when the electrostatic potential decays to its bulk value 
with the opposite sign. 
These behaviors are well explained by a simple mathematical form
of the electro-osmotic coefficient.
\end{abstract}


\maketitle
\section{Introduction}

Aqueous solutions of polyelectrolytes are widely seen in many 
systems,\cite{Rubinstein_Papoian_2012} 
and are used in a variety of industrial applications. 
Some bio-macromolecules, such as proteins and DNA, 
also exist in the charged state. 
The chemical and physical properties 
of polyelectrolyte solutions have been extensively studied.
\cite{deGennes_1979,Oosawa_1971,Barrat_Joanny_1996,
Joanny_Leibler_1990,Dobrynin_Rubinstein_2005}
When a polyelectrolyte solution flows near a charged interface, 
electrokinetic phenomena occur. 
The electrokinetics of polyelectrolyte solutions 
are important for understanding some physiological situations 
such as strain and restoration of bone\cite{Igarashi_1991} and 
blood flows in capillary vessels.\cite{Liu_Yang_2009}
However, although electro-osmosis and electrophoresis have been well 
investigated in simple electrolyte solutions,\cite{O'Brien_White_1978} 
the electrokinetics 
in polyelectrolyte solutions are not entirely understood.
\cite{Burgreen_Nakache_1964,Rice_Whitehead_1965,Levine_Marriott_Robinson_1974}

The static properties 
of electrically neutral polymers near 
solid walls have been described by mean-field and scaling theories.
\cite{Joanny_Leibler_deGennes_1979,deGennes_1981,deGennes_1982} 
According to prediction,
the spatial decay of the concentration profile 
from the surface 
to the bulk follows an
exponential or power law. 
If the chemical intermolecular interactions 
between the polymer and the wall surface are sufficiently strong, 
adsorption or depletion layers form on the wall. 
By contrast, in polyelectrolyte solutions, 
the concentration profile and conformations of 
the polymer chains depend on 
the interplay between the electrostatic and 
chemical interactions. 
\cite{Shafir_Andelman_2004,Shafir_Andelman_Netz_2003,Joanny_1999,
Borukhov_Andelman_Orland_1998,Chatellier_Joanny_1996,
Borukhov_Andelman_Orland_1995} 
Electrostatic interactions 
attract the  polyelectrolytes to the oppositely charged wall,
and {\it vice versa} thereby, lead to formation of the adsorption layer.
If the polyelectrolyte adsorption is strong, 
the electrostatic potential is highly modified 
and the concentration profile becomes nonmonotonic. 
In the phenomenon known as {\it charge inversion} 
or {\it overcharging}, the electrostatic potential 
decays to its bulk value with reversed sign.

A few experimental and theoretical studies on 
the electrokinetics in polyelectrolyte solutions 
have been reported. 
The first measurements of electro-osmosis in a polyelectrolyte 
solution did not account for the near-surface structures.\cite{Vink_1988}
Later researchers noted the importance of fluid viscosity in the electric 
double layer.\cite{Bello_1994, Otevrel_Kleparnik_2002}
Electro-osmosis is also thought to be influenced by the non-Newtonian 
behavior of polymeric liquids.\cite{Zhao_Zholkovskij_Masliyah_Yang_2008,
Barrat_Joanny_1996,Olivares_VeraCandioti_Berli_2009,Zhao_Yang_2011} 
However, instead of considering electrostatic and chemical interactions, 
these studies assumed a given static profile near the surface. 
This study aims to clarify the electro-osmosis of 
polyelectrolyte solutions, focusing on 
the structure of electric double layers. 
In determining the electro-osmotic coefficient, 
both viscosity near the interface and 
the strength of the overcharging are considered.

\section{Macroscopic electro-osmosis} 

When a charged-surface capillary is 
filled with an electrolyte solution 
and subjected to an electric field $E$ and 
a pressure difference $P(=-\nabla p)$, 
a volume flux and electric current are induced 
along the external fields. 
Here $p$ is the pressure. 
When $E$ and $P$ are sufficiently weak, 
the volume flux and the electric current 
are given by 
\begin{eqnarray}
V&=&L_{11}P+L_{12}E,
\label{eq:Onsager1}\\
J&=&L_{21}P+L_{22}E. 
\label{eq:Onsager2}
\end{eqnarray}
$V$ and $J$ are the mean volume flux and mean electric current, 
respectively, and 
$L_{ij}$s are the Onsager transport coefficients. 
This article focuses on $L_{12}$, 
known as the {\it electro-osmotic coefficient}. 
In a solution containing many monovalent ions, 
the electro-osmotic flow induced by the electric field is non-Poiseuille. 
Instead, it presents as a plug flow that exponentially decays with 
the Debye screening length. 
When the capillary diameter is much larger than the Debye screening 
length, 
the electro-osmotic coefficient is independent of the salt 
concentration, and is  given by 
Smoluchovski's formula, 
$L_{12}=-\varepsilon\psi_\mathrm{S}/4\pi\eta$, 
in which $\varepsilon$ is the dielectric constant of the solution, 
$\psi_\mathrm{S}$ is the electrostatic potential at the nonslip surfaces, 
and $\eta$ is the viscosity of the solution.
The near-surface properties of polyelectrolyte solutions in capillaries 
greatly differ from those of small-ion electrolyte solutions. 
For example, as mentioned above, polyelectrolyte solutions form
adsorption or depletion layers near the capillary surface. 
 
Also, electro-osmosis in polymer solutions is likely to be affected by 
the rheology of the solutions.
In particular, the viscosity of such solutions is not constant but 
depends on the local polymer concentration and the shear rate. 
Later in this study, we account for both of these factors in determining
the electro-osmotic coefficient using mean-field 
equations.

\section{mean-field equations for concentrations, 
electrostatic potential and flow} 

\begin{figure}
\includegraphics[width=0.5\textwidth]{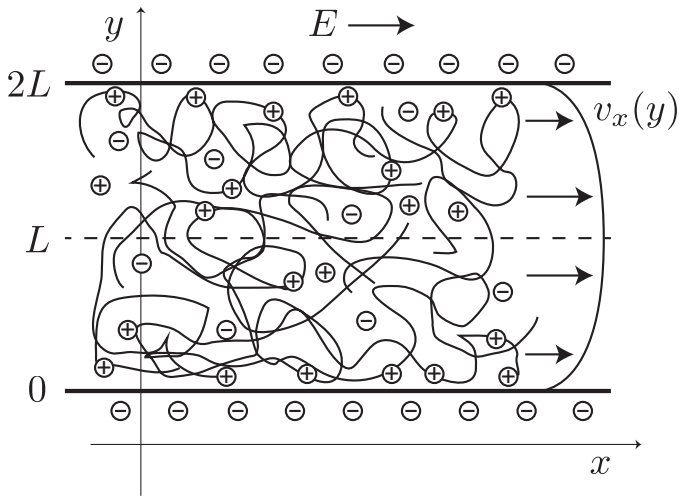}
\caption{Schematic of the investigated system. 
A positively charged polyelectrolyte solution
is confined within a negatively charged slit and an external electric field 
is applied along the slit walls. The long-chain polymers are 
interspersed with polymer counterions and anions derived from salt.}
\label{fig1}
\end{figure}

We consider an aqueous solution 
of sufficiently long polyelectrolyte chains in 
a slit (see Fig.~\ref{fig1}).
A fraction $f$ of the polyelectrolytes is positively charged,
whereas the slit wall is negatively charged. 
Counterions from the polyelectrolytes 
and salts are also dissolved in the solution. 
For simplicity, we assume that the anions from the salt and the counterions 
from the polymers are the same species and all the small ions are 
monovalent.
The free energy of the system is contributed by polymer conformations, 
ion distributions, and electrostatic interactions as follows:  
\begin{equation}
F=F_\mathrm{poly}+F_\mathrm{ions}+F_\mathrm{ele}.
\end{equation}
The polymer free energy is given by\cite{deGennes_1979} 
\begin{equation}
F_\mathrm{poly}
=k_\mathrm{B}T\int d\boldsymbol{r}
\left[\frac{a^2}{6}|\nabla\phi|^2+\frac{v}{2}\phi^4\right],  
\end{equation}
where $\phi$ is an order parameter related to the 
local polymer concentration $c(\boldsymbol{r})$, 
given by $\phi(\boldsymbol{r})=\sqrt{c(\boldsymbol{r})}$. 
$k_\mathrm{B}T$ is the thermal energy, $a$ is the monomer size, 
and $v$ is the second virial (excluded volume) coefficient of the monomers.

The ion free energy, contributed by 
the translational entropy of the ions, is given by
\begin{equation}
F_\mathrm{ion}=k_\mathrm{B}T
\int d\boldsymbol{r}\sum_{i=\pm}\left[c^i\ln (c^ia^3)-c^i\right],
\end{equation}
where $c^+(\boldsymbol{r})$ 
and $c^-(\boldsymbol{r})$ 
are the concentrations of the cations and anions, respectively. 
The electrostatic free energy is given by
\begin{equation}
F_\mathrm{ele}=\int d\boldsymbol{r}
\left[\rho\psi-\frac{\varepsilon}{8\pi}|\nabla\psi|^2\right]. 
\label{eq:F_ele} 
\end{equation}
where $\psi(\boldsymbol{r})$ is the local electrostatic potential, 
$\varepsilon$ is the dielectric constant of the aqueous solution, 
and $\rho(\boldsymbol{r})$ is the electric charge density defined as 
\begin{equation}
\rho=e(fc+c^+-c^-),
\end{equation}
where $e$ is the elementary electric charge. 

The control parameters in this study are the bulk concentrations of the  
cation $c^+_\mathrm{b}$ and the charged monomer fraction $f$. 
These parameters should satisfy the neutral charge condition, 
$fc_\mathrm{b}+c^+_\mathrm{b}-c^-_\mathrm{b}=0$, 
in the bulk. 
Here $c_\mathrm{b}$ and $c_\mathrm{b}^{-}$ are the 
bulk concentrations of the monomers and anions, respectively, whose
steady profiles are obtained by minimizing the following grand potential
\begin{equation}
\Xi=F-\mu\int \phi^2d\boldsymbol{r}-\sum_{i=\pm}\mu^i\int c^id\boldsymbol{r},
\end{equation}
where $\mu$ and $\mu^i$ ($i=\pm$) 
denote the chemical potential of each component.

The solution is confined within 
a slit bounded by two parallel walls. 
We assume that 
the above variables change only along the $y$ axis, 
and are homogeneous along the $x$ and $z$ axes. 
In this scenario, the
mean-field equations are 
\begin{eqnarray}
\frac{a^2}{6}\frac{\partial^2\phi}{\partial y^2}
&=&v(\phi^3-c_\mathrm{b}\phi)+f\phi\beta e\psi,
\label{eq:Edwards}\\
\frac{\partial^2\psi}{\partial y^2}&=&-\frac{4\pi e}{\varepsilon}
f(\phi^2-c_\mathrm{b}\exp[\beta e\psi])
+\frac{8\pi e}{\varepsilon}c_\mathrm{b}^+\sinh(\beta e\psi),\nonumber 
\label{eq:PB}\\
\end{eqnarray}
where $\beta=1/k_\mathrm{B}T$. 
Eq.~(\ref{eq:Edwards}) is the Edwards equation that accounts for
the charge effect, while
Eq.~(\ref{eq:PB}) is the Poisson-Boltzmann equation for the system 
containing the salts and polyelectrolytes. 

Applying a sufficiently weak electric field $E$ 
in the $x$ direction, the system evolves to steady state 
in which ion fluxes are induced along $E$. 
Because $E$ is weak and orthogonal to $-\nabla \psi(y)$, 
we assume that it influences neither 
the concentration fields nor the polymer conformations 
(see Appendix \ref{app:A}). 
In steady state, 
the mechanical forces are balanced. 
This force balance is expressed by the Navier-Stokes equation, 
whose simplified form is 
\begin{equation}
\frac{\partial }{\partial y}\left[\eta(\phi)
\frac{\partial v_x}{\partial y}\right]+\rho E=0,
\label{eq:NS}
\end{equation}
where $v_x(y)$ is the $x$ component of the velocity field. 
In this case, because we impose no pressure difference on the system, 
$P=0$ in Eqs.~(\ref{eq:Onsager1}) and (\ref{eq:Onsager2}). 
$\eta(\phi)$ is the viscosity, which is a function of 
the concentration order parameter $\phi$. 
In this study, we set 
\begin{eqnarray}
\eta(\phi)=\eta_0\left\{ 1+h (\phi/\sqrt{c_{\rm b}})^\alpha\right\}, 
\end{eqnarray}
where $h$ and $\alpha$  are nondimensional parameters.
Here $\eta_0$ is the solvent viscosity and 
$\eta_{\rm b}=\eta_0(1+h)$ denotes the viscosity in the bulk. 
Because $\eta(\phi)$ usually increases from $\eta_0$ 
as $\phi$ increases, $h$ and $\alpha$ are assumed positive. 
As described in Appendix \ref{app:B}, 
$h$ and $\alpha$ depend on 
the physical parameters $N$, $f$, and $c_{\rm b}^+$, 
in which $N$ is the polymer length. 
In this study, however, $h$ is assumed as an independent parameter. 
According to Fuoss law \cite{Dobrynin_Colby_Rubinstein_1995}, 
we set $\alpha=1$. 
Later, we demonstrate that
these simplifications do not alter the essential results. 

As shown in Fig.~\ref{fig1}, 
the surfaces are placed 
at $y=0$ and $2L$, where $2L$ is the slit width 
and the electrostatic potentials are the same at both surfaces. 
Because all profiles are symmetric with respect to $y=L$, 
we consider only the range $[0,L]$. 
At the bottom surface $(y=0)$, 
we assume $\phi(0)=0$, implying that
the intermolecular interactions between the surfaces and polymers 
are strongly repulsive. 
We also set 
$v_x(0)=0$ and $\psi(0)=\psi_\mathrm{S}$. 
The former is the nonslip boundary condition for the flow. 
$\psi_\mathrm{S}$ is negative because the surfaces are negatively 
charged and the electrostatic interaction 
between the polymer and surfaces is attractive. 
Because the system is symmetric, 
all $y$ derivatives vanish at $y=L$; 
\begin{eqnarray}
\left.\frac{\partial \phi}{\partial y}\right|_{y=L}=0, 
\left.\frac{\partial \psi}{\partial y}\right|_{y=L}=0, 
\left.\frac{\partial v_x}{\partial y}\right|_{y=L}=0. 
\label{eq:BC} 
\end{eqnarray}

In this study, we assume that the electric field is sufficiently weak 
so that the flow speed is proportional to the field strength. 
Specifying a coefficient $\lambda_{12}(y)$, 
the flow profile is expressed as $v_x(y)=\lambda_{12}(y)E$. 
In other words, 
the solution is Newtonian and 
the nonlinear dependence of 
the flow on $E$ can be ignored. 
Having obtained the static profiles 
(which are difficult to solve analytically), 
$\lambda_{12}$ is calculated as 
\begin{equation}
\lambda_{12}(y)=\frac{\varepsilon}{4\pi}
\int^y_0\frac{dy'}{\eta(\phi(y'))}\left.
\frac{\partial \psi}{\partial y}\right|_{y'}.
\label{eq:lambda12}
\end{equation} 
This quantity is related to the macroscopic electro-osmotic coefficient 
in Eq.~(\ref{eq:Onsager1}) by 
\begin{equation}
L_{12}=\frac{1}{L}\int^L_0\lambda_{12}(y)dy.
\label{eq:L12}
\end{equation}

\section{Results and Discussion}

To study the effects of the near-surface polyelectrolyte structures 
on electro-osmosis in this system, we numerically evaluate 
Eqs.~(\ref{eq:Edwards}), 
(\ref{eq:PB}) and (\ref{eq:NS}). 
The parameter settings are $c_\mathrm{b}=10^{-6 }$\AA$^{-3}$, 
$v=50 $\AA$^3$, 
$L=1024 $\AA, 
$\ell_\mathrm{B}=7 $\AA, 
$T=300 $K, 
$\psi_\mathrm{S}=-k_\mathrm{B}T/e=-25.8 $mV, 
$a=5 $\AA, and 
$\eta_0=0.01 $P. 
Here $\ell_\mathrm{B}$ is the Bjerrum length, given by
$\ell_\mathrm{B}=e^2/(\varepsilon k_\mathrm{B}T)$. 
The polymer chains are assumed so long that $c_\mathrm{b}>c^*$, 
where $c^*$ is the overlap concentration of the polymer solution 
(see Appendix \ref{app:B}).  

The electro-osmotic coefficient is evaluated from the $L_{12}^0$ 
of a solution without polyelectrolytes, 
given by 
$L_{12}^0=-\varepsilon\psi_\mathrm{S}/4\pi\eta_0=
1.82\times 10^{-4} $cm$^2$/V$\cdot$s.
The space discretization in the numerical calculations 
is $d=1 $\AA.

\subsection{Electrically neutral polymer solution with chemically 
repulsive surfaces}

First, we assume that polymers are electrically neutral, 
{\it i.e.}, $f=0$. 
In this case, the mean-field equations (\ref{eq:Edwards}) and 
(\ref{eq:PB}) are exactly solved as 
\begin{eqnarray}
\phi&=&\sqrt{c_\mathrm{b}}\tanh\left(\frac{y}{\xi}\right),
\label{eq:phi_f=0}\\
\beta e\psi&=&2\ln\frac{1+\mathrm{e}^{-\kappa y}
\tanh(\beta e\psi_\mathrm{S}/4)}{1-\mathrm{e}^{-\kappa y}
\tanh(\beta e\psi_\mathrm{S}/4)}.
\label{eq:psi_f=0} 
\end{eqnarray}
where $\kappa
=(8\pi\ell_\mathrm{B}c_\mathrm{b}^+)^{1/2}$ is 
the Debye wave number 
and  $\xi=a/\sqrt{3vc_\mathrm{b}}=408 $\AA\, 
is the correlation length of the polymer concentration fluctuation. 
$\mathrm{e}$ is Napier's constant. 
Note that these analytical solutions are valid only 
when $\kappa^{-1}\ll L$ and $\xi\ll L$ 
because 
they are solved under the boundary conditions at $y=0$ and $L$. 
If $|\beta e\psi_\mathrm{S}|\ll 1$, 
Eq.~(\ref{eq:psi_f=0}) reduces to 
\begin{equation}
\psi=\psi_\mathrm{S}\mathrm{e}^{-\kappa y}, 
\label{eq:DebyeHuckel}
\end{equation}
using the Debye-H\"{u}ckel approximation. 

If the slit width is much larger than all other length scales in the system, 
$L_{12}$ is approximately equal to $L_{12}\approx \lambda_{12}(L)$. 
Therefore, we write 
\begin{equation}
L_{12}\approx \frac{\varepsilon}{4\pi\eta_0}\int^L_0
\frac{dy}{1+\eta_1\phi}\frac{\partial \psi}{\partial y}.
\label{eq:L12_f=0} 
\end{equation}
Because $\psi(y)$ is a monotonically increasing function 
of $y$ in Eqs.~(\ref{eq:psi_f=0}) and (\ref{eq:DebyeHuckel}), 
the integral $\int \cdots dy$ in Eq.~(\ref{eq:L12_f=0}) can be replaced 
by $\int \cdots d\psi$, using $\mathrm{e}^{y/\xi}=\zeta^{-1/\kappa\xi}$.
$L_{12}$ is then calculated as 
\begin{eqnarray}
\frac{L_{12}}{L_{12}^0}=
\int^1_0 \frac{\left(\zeta^{-1/\kappa\xi}
+\zeta^{1/\kappa\xi}\right)d\zeta}
{(h+1)
\zeta^{-1/\kappa\xi}-(h-1)
\zeta^{1/\kappa\xi}},\nonumber\\
\label{eq19}
\end{eqnarray}
where $\zeta=\psi/\psi_\mathrm{S}$ is a reduced 
electrostatic potential. 
After some calculations, Eq.~(\ref{eq19}) can be expanded as 
\begin{eqnarray}
&&\frac{L_{12}}{L_{12}^0}
=\frac{2}{h+1}\frac{\kappa\xi+1}{\kappa\xi+2}+\frac{1}{h+1}\sum_{n=1}^\infty 
\left(\frac{h-1}{h+1}\right)^n
\nonumber\\
&&\times \left[\frac{1}{2n/(\kappa\xi)+1}+
\frac{1}{2(n+1)/(\kappa\xi)+1}\right].
\label{eq:series}
\end{eqnarray}
When $h=1$, Eq.~(21) reduces to
\begin{equation}
L_{12}=L_{12}^0\left(\frac{\kappa\xi+1}
{\kappa\xi+2}\right). 
\end{equation}
Clearly, Eq.~(22) is an increasing function of $\kappa\xi$. 

\begin{figure}
\includegraphics[width=0.3\textwidth]{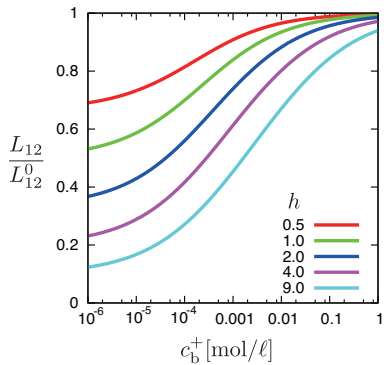}
\caption{
Electro-osmotic coefficients $L_{12}$ 
in the solution without polyelectrolytes, plotted 
as functions of salt concentration $c_\mathrm{b}^+$. 
In place of polyelectrolytes, 
electrically neutral polymers are dissolved. 
The viscosity parameter $h$ is varied. 
}
\label{fig2}
\end{figure}

The electro-osmotic coefficient 
calculated by Eq.~(\ref{eq:series}) is plotted as a function of 
salt concentration in Fig.~\ref{fig2}. 
Shown are the coefficients 
for several values of the bulk viscosity parameter $h$. 
As the salt concentration increases, 
the electro-osmotic coefficient increases and 
approaches $L_{12}^0$, regardless of $h$. 
By contrast, in the low salt concentration regime, 
$L_{12}$ decreases as $(c_\mathrm{b}^+)^{1/2}$ to 
$L_{\rm 12}^\mathrm{b}=L_{12}^0/(1+h)$, 
the electro-osmotic coefficient estimated at 
the viscosity of the bulk solution. 

We interpret these results as follows. 
In the neutral polymer solution, 
the electrostatic interaction does not 
influence the polymer concentration profile. 
The polymers are depleted from the surface 
by short-ranged surface forces, 
and the near-surface viscosity is smaller than that in the bulk. 
Only the region near the surface, where $\rho\ne 0$, 
responds to the applied electric field. 
The charged region is characterized by the Debye length 
$\kappa^{-1}$ from the surface. 
If the Debye length is smaller than the correlation length 
$\xi$, the electro-osmosis is enhanced;
otherwise it is suppressed.

Given the effective viscosity $\eta_{\rm S}$, 
the electro-osmotic coefficient is 
calculated by the usual Smoluchowski's 
formula, $L_{12}=-\varepsilon\psi_\mathrm{S}/4\pi\eta_\mathrm{S}$.
As noted above, the formation of the depletion layer near the surface 
effectively lowers the viscosity of the solution. 
From Eq.~(\ref{eq:series}), 
the effective viscosity decreases with $\kappa\xi$ 
as 
\begin{eqnarray}
\eta_\mathrm{S}
\approx \eta_0(1+h)\left\{
1-\kappa\xi\frac{h}
{h-1}\ln \frac{h+1}{2}
\right\},\nonumber\\
\label{eq:eta_s} 
\end{eqnarray}
when $\kappa\xi\ll 1$, using $\sum_{n=1}^\infty n^{-1}r^n=\ln[1/(1-r)]$. 
On the other hand, when $\kappa\xi\gg 1$, 
the viscosity approaches the solvent viscosity, 
$\eta_\mathrm{S}\approx \eta_0$. 
This phenomenon can be explained as follows. 
In the high salt limit, the electrostatic interaction between ions and 
walls is screened by a short length scale.
If the wall is chemically repulsive to the polymers, 
the polymers are depleted from the surface with a
correlation length far exceeding the Debye screening length. 

\subsection{Polyelectrolyte solution with electrically 
attractive and chemically repulsive surfaces}

\begin{figure}
\includegraphics[width=0.5\textwidth]{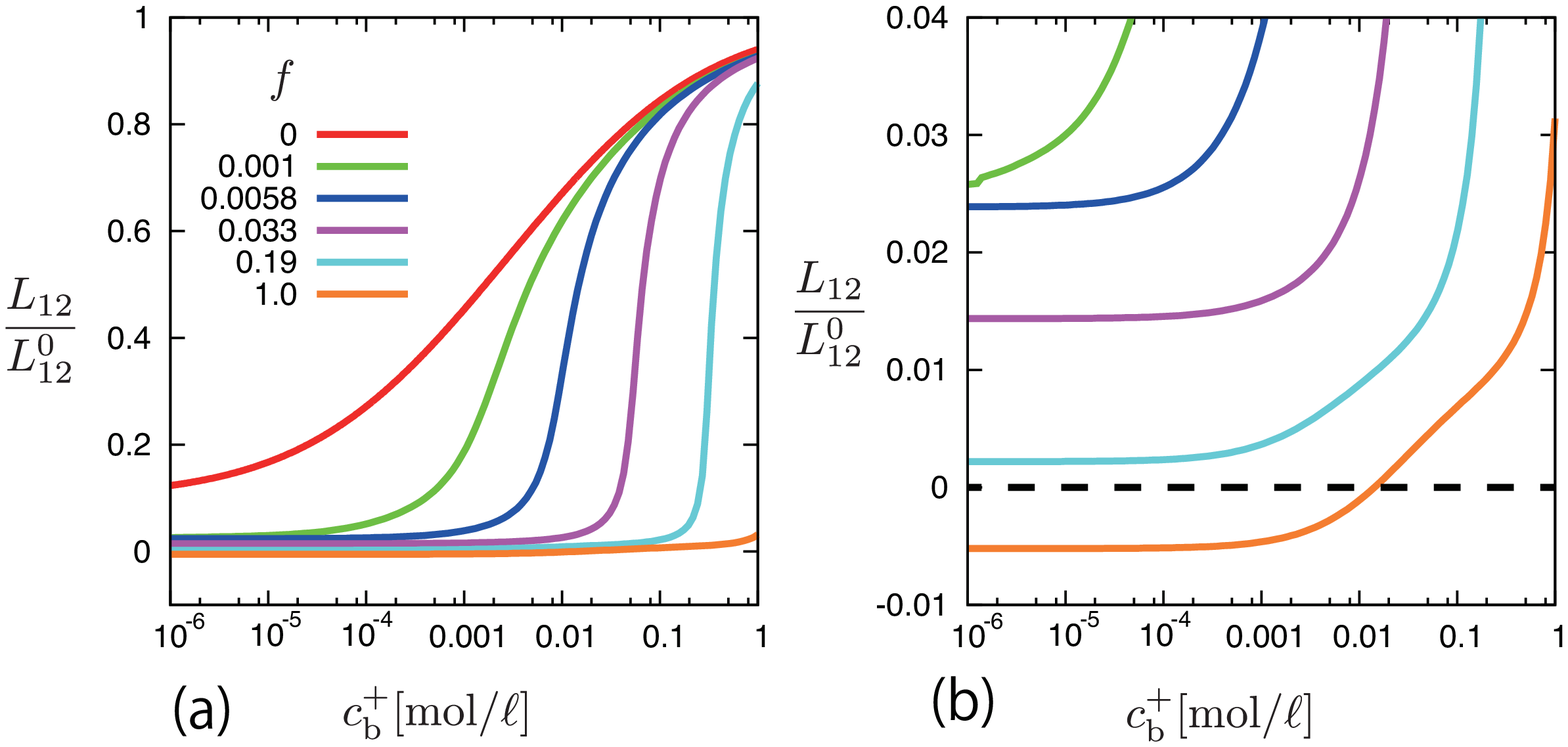}
\caption{
(a) Electro-osmotic coefficients $L_{12}$, plotted  
as functions of salt concentration $c^+_\mathrm{b}$. 
The fraction of charged monomers in the polyelectrolyte $f$ is varied for 
fixed $h=9$. 
At $f = 0$, the curve is that of the electrically neutral polymer solution, 
and it increases with $c^+_\mathrm{b}$ 
as shown in Fig.~\ref{fig2}. 
At sufficient salt concentrations, all curves approach $L^0_{12}$. 
 (b) Magnification of the same plots around a small range of $L_{12}$. 
}
\label{fig3}
\end{figure}

Next, we consider polyelectrolyte solutions, {\it i.e.}, $f\ne 0$. 
Figure \ref{fig3}(a) shows the 
electro-osmotic coefficients as functions of salt concentration. 
Here we fix $h=9$ and vary the fraction of 
charged monomers $f$. 
We find that, as in neutral polymer solutions (see Fig.~\ref{fig2}),
electro-osmosis is suppressed in the low salt regime. 
At high salt concentrations, 
the electro-osmotic coefficient approaches 
$L_{12}^0$. 
Figure \ref{fig3}(a) also indicates that, 
with increasing electric charge on the polyelectrolytes,
electro-osmosis becomes more suppressed 
and salinity exerts a more drastic 
effect. 
In Fig.~\ref{fig3}(b), these plots are magnified  around $L_{12}=0$. 
Interestingly, 
the electro-osmotic coefficient can become negative 
at sufficiently dilute salt and 
when the polyelectrolytes are highly charged. 
Such inversion of electro-osmotic flow is never 
observed in neutral polymer solutions.

\begin{figure}
\includegraphics[width=0.5\textwidth]{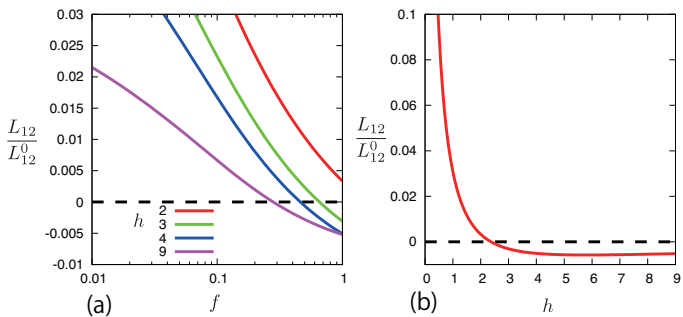}
\caption{
(a) Electro-osmotic coefficients $L_{12}$, plotted 
as functions of the fraction of charged monomers 
in the polyelectrolytes $f$. 
$h$ is varied at fixed salt concentration $c_\mathrm{b}^+=10^{-6} $mol/$\ell$. 
 (b) Electro-osmotic coefficient $L_{12}$ plotted
as a function of $h$. 
We set $c_\mathrm{b}^+=10^{-6} $mol/$\ell$ and $f=1$.
}
\label{fig4}
\end{figure}

The electo-osmotic coefficients are plotted
as functions of $f$ in Fig.~\ref{fig4}(a). 
Here the salt concentration is fixed at a 
low concentration $c_\mathrm{b}^+=10^{-6}$[mol/$\ell$], 
and the bulk viscosity parameter $h$ is changed. 
We observe that the electro-osmotic flow is weakened 
if the polyelectrolytes are 
highly charged. 
The mechanism of this phenomenon will be discussed later. 
Figure~\ref{fig4}(a) also shows that electro-osmosis inversion occurs 
only at sufficiently high $h$. 
Figure~\ref{fig4}(b) plots the electro-osmotic coefficient 
versus $h$ for 
$c_\mathrm{b}^+=10^{-6}$[mol/$\ell$] and $f=1$. 
As discussed above, 
the electro-osmotic flow in neutral polymer solutions 
saturates at $L_{12}^{\rm b}$ in the 
low salinity limit, according to Eq.~(\ref{eq:eta_s}). 
However, this equation cannot explain the curve in Fig.~\ref{fig4}(b).

\subsubsection{Relationship between electro-osmosis and static properties}

\begin{figure}
\includegraphics[width=0.5\textwidth]{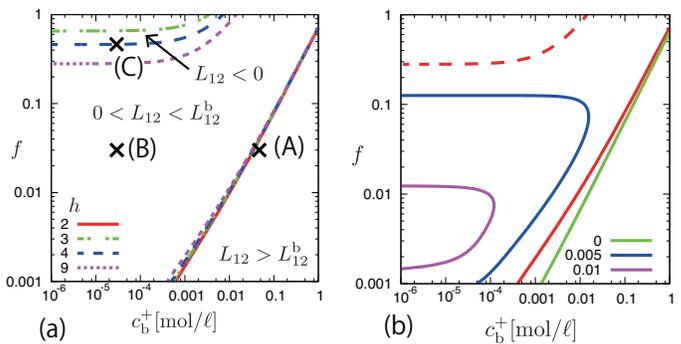}
\caption{
(a) State diagram of the electro-osmotic flow in 
the $c^+_\mathrm{b}$-$f$ plane. 
Because the polyelectrolyte concentration varies 
in the slit, the local viscosity depends on the distance from the wall. 
$L^\mathrm{b}_{12}$ is the electro-osmotic coefficient, estimated 
from the shear viscosity of the bulk solution. 
(b) Contours of the amount of excess adsorption, $\Gamma$, 
for $\Gamma=0,\;0.005$ and $0.01$. 
Shown are the contour lines of $L_{12}=0$ (solid) 
and $L_{12}=L_{12}^{\rm b}$ (broken) 
for $h=9$. 
The $\Gamma=0$ contour behaves similarly to 
the line $L_{12}=L_{12}^\mathrm{b}$ in (a).}
\label{fig5}
\end{figure}

Figure.~\ref{fig5}(a) plots the curves of $L_{12}=0$ and 
$L_{12}=L_{12}^\mathrm{b}$ 
in a $c_\mathrm{b}^+$-$f$ plane. 
As the bulk viscosity parameter $h$ decreases, 
the region of inverted electro-osmosis ($L_{12}<0$) 
shrinks and eventually disappears as
$h$ becomes small.  
On the other hand, 
the $L_{12}=L_{12}^\mathrm{b}$ curves are less sensitive to changes in $h$. 
This implies that $L_{12}$ around $L_{12}^\mathrm{b}$ 
depends more on the static than kinetic
properties.

\begin{figure}
\includegraphics[width=0.5\textwidth]{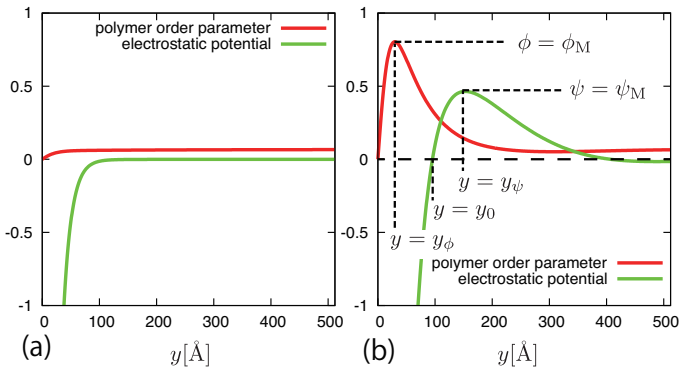}
\caption{
Profiles of the polymer order parameter $\phi$ 
and the electrostatic potential $\psi$ 
near the surface. 
The bulk concentrations of the salt 
are (a) $c_\mathrm{b}^+=0.0476$[mol/$\ell$]
and (b) $c_\mathrm{b}^+=2.91\times 10^{-5}$[mol/$\ell$].
In both cases, the fraction of charged monomer in the 
polyelectrolyte is $f=0.03$.
The profiles in (a) and (b) correspond to conditions (A) 
and (B) in Fig.~\ref{fig5}(a). 
For a clearer representation, we 
plot $\phi(y)/15\phi_\mathrm{b}$ and $15\psi(y)/|\psi_\mathrm{S}|$ 
rather than $\phi(y)$ and $\psi(y)$. 
}
\label{fig6}
\end{figure}

To characterize the static properties, we define a quantity $\Gamma$ as 
\begin{equation}
\Gamma=\int^L_0dy (c-c_\mathrm{b}).
\end{equation}
$\Gamma$ measures the amount of excess adsorption of the polyelectrolytes. 
Figure~\ref{fig5}(b) plots the contour lines of $\Gamma$ in 
the $c_\mathrm{b}^+$-$f$ plane. 
The $\Gamma=0$ contour characterizes the adsorption-depletion 
transition.\cite{Shafir_Andelman_Netz_2003}
In the system investigated here, 
the positively charged polymers are dissolved in the slit 
between the negatively charged walls. 
Electrostatic interaction adheres the
polymers to the oppositely charged wall surface. 
On the other hand, intermolecular interaction prevents 
the polymers from directly contacting 
the surface (see Fig.~\ref{fig6}(a)). 
When the electrostatic interaction is well screened by high salt content, 
chemical interaction depletes the polymers from the 
surface vicinity. 

Interestingly, when $c_\mathrm{b}^+$ is fixed, 
excess adsorption does not continuously increase toward $f=1$
but instead peaks at an intermediate $f$. 
As shown in Fig. ~\ref{fig5}(b), the polyelectrolytes 
with $c_\mathrm{b}^+=10^{-6} $mol/$\ell$ 
are most strongly adsorbed when $f\approx 5\times 10^{-3}$. 
This nonmonotonic behavior is counterintuitive because 
one expects that highly charged polyelectrolytes 
will be adsorbed with greatest strength. 
The adsorption-depletion transition has been intensively
studied by Shafir {\it et al.}\cite{Shafir_Andelman_Netz_2003}
Comparing Fig.~\ref{fig5}(a) and (b), 
we find that the curves $L_{12}=L_{12}^\mathrm{b}$ roughly 
coincide with that of $\Gamma=0$. 
When the polymers are adsorbed to the surface ($\Gamma>0$), 
the electro-osmotic coefficient is smaller than that 
determined by the surface potential and bulk viscosity 
$L_{12}^\mathrm{b}$, and {\it vice versa}. 

\begin{figure}
\includegraphics[width=0.5\textwidth]{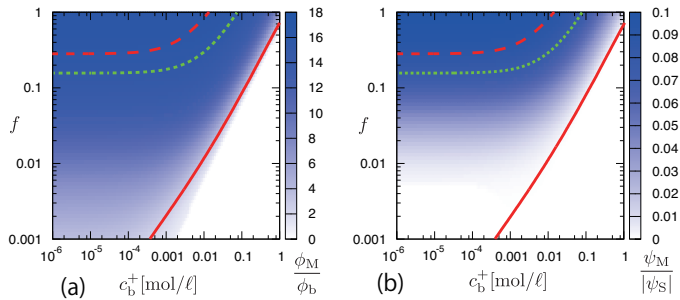}
\caption{
Effect of $c_{\rm b}^+$ and $f$ on the peak heights of 
(a) concentration profile 
$\phi_{\rm M}$ and (b) electrostatic potential $\psi_{\rm M}$. 
Shown are the contour lines of $L_{12}=0$ (broken) and 
$L_{12}=L_{12}^{\rm b}$ (solid) 
for $h=9$. 
The dotted line is $L_{12}=0$ estimated by Eq.~(\ref{eq:L12_est}). 
Uncolored regions indicate where no peaks appear in $\phi$ and 
$\psi$ 
({\it i.e.}, where $\phi_\mathrm{M}$ and $\psi_\mathrm{M}$ are undefined). 
}
\label{fig7}
\end{figure}

Figure \ref{fig6} shows profiles of the 
polymer concentration and electrostatic potential 
at (a) high and (b) low salt concentrations. 
The fraction of charged monomers is $f=0.03$. 
The solution conditions 
are as indicated in Fig.~\ref{fig5}(a). 
Under low-salinity conditions, where $L_{12}<L_{12}^\mathrm{b}$,  
a peak appears in the concentration profile.
Hereafter, the height and the position of the peak are 
denoted as  $\phi_{\rm M}$ and $y_\phi$, respectively. 
As shown in Fig.~\ref{fig5}(b), the amount of 
adsorption is positive ({\it i.e.}, in excess) 
Hence, we refer to the region of $\phi>\phi_\mathrm{b}(=\sqrt{c_{\rm b}})$ as 
an adsorption layer, although the polymers 
themselves do not contact the surface. 
The electrostatic potential also peaks 
at $y=y_{\rm \psi}$. 
We call this peak  
an {\it overcharging potential} and its height 
is denoted as $\psi_\mathrm{M}$. 
We should note that the $\phi$ and $\psi$ peaks appear
at different positions, with $y_{\psi}>y_\phi$. 
We also define $y_0$, which satisfies $\psi(y_0)=0$. 
As discussed below, 
the adsorption layer and the 
overcharging potential play essential roles in 
the decrease and 
inversion of the electro-osmotic coefficient. 
Conversely, under high-salinity conditions, 
the profiles monotonically increase 
to the bulk values without developing peaks. 
The dependences of $\phi_{\rm M}$ and $\psi_{\rm M}$ on 
$c_{\rm b}^+$ and $f$ are shown in Figs.~\ref{fig7}(a) and (b), 
respectively. 
The contours of $\phi_{\rm M}$ and $\psi_{\rm M}$ are 
qualitatively similar to that of $L_{12}$ in Fig.~\ref{fig5}(a) 
but are dissimilar from that of the 
excess adsorption. 
This implies that the maximum amount of adsorption 
is not important in the electro-osmotic phenomena.

The uncolored region, in which the profile does not peak, 
almost coincides with that of $L_{12}>L_{12}^{\rm b}$. 
The gradient of 
the electro-osmotic flow is localized to the range of the Debye screening 
length from the surface (see Fig.~\ref{fig8}(a)). 
Therefore, the formation of the depletion layer effectively 
reduces the solvent viscosity. 
Because the electro-osmotic flow is inversely proportional to 
the viscosity, 
depletion enhances the electro-osmosis. 
At the adsorption-depletion transition,
the increase in $L_{12}$ caused by 
the depletion cancels 
the decrease caused by adsorption.  
Then, the $\Gamma=0$ curve is roughly consistent 
with that of $L_{12}=L_{12}^{\rm b}$. 
Because neutral polymers in solution do not adhere to the surface, 
electro-osmosis is more strongly suppressed in polyelectrolyte solutions 
than in neutral polymer solutions. 

The uncolored region in Fig.~\ref{fig7}(b), where 
$\psi_{\rm M}$ develops no peak, 
is slightly wider than in Fig.~\ref{fig7}(a),
where $\phi_{\rm M}$ develops no peak. 
This difference is delicate because the Debye screening length 
becomes comparable to the system size when $c_{\rm b}^+$ and $f$ 
are very small. 

\subsubsection{Relationship between electro-osmosis and dynamical properties}
\begin{figure}
\includegraphics[width=0.5\textwidth]{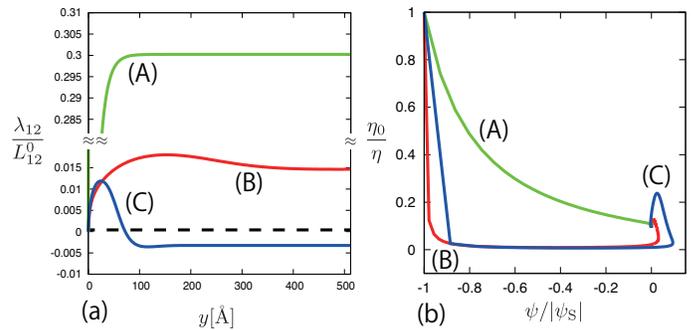}
\caption{
(a) Profiles of the electro-osmotic flow $\lambda_{12}$ 
in three states; 
(A) depletion state: $c_\mathrm{b}^+=0.0476 $mol/$\ell$ and $f=0.03$. 
(B) adsorption state: $c_\mathrm{b}^+=2.91\times 10^{-5} $mol/$\ell$ 
and $f=0.03$.
(C) flow inversion state: $c_\mathrm{b}^+=2.91\times 10^{-5} $mol/$\ell$, 
and $f=0.46$. 
These conditions are marked in Fig.~\ref{fig5}(a). 
In condition (C), the overcharging potential is 
$\psi_\mathrm{M}/|\psi_\mathrm{S}|\approx 0.1$.
(b) Parametric representations of $\psi(y)$ and $1/\eta(\phi(y))$ 
for the three states. 
Points $(\psi/|\psi_{\rm S}|,\eta_0/\eta)=(-1,1)$ and 
$(\psi/|\psi_{\rm S}|,\eta_0/\eta)=(0,1/(1+h))$ correspond to 
the surface ($y=0$) and bulk ($y=L$), respectively. 
The bulk viscosity parameter is fixed at $h=9$.}
\label{fig8}
\end{figure}

As shown in Fig.~\ref{fig7}, $\phi_{\rm M}$ and $\psi_{\rm M}$ are 
large in the regime of large $f$ and small $c_{\rm b}^+$, where
the electro-osmotic coefficient becomes negative. 
We emphasize that 
these large values of $\phi_{\rm M}$ and $\psi_{\rm M}$ are 
essentially important for the sign reversal of $L_{12}$. 

Figure \ref{fig8}(a) shows the profiles of the flow field near the surface 
under three conditions. 
Here we note that $v_x(y)=\lambda_{12}(y) E$. 
Conditions (A) and (B) correspond to 
the adsorption and depletion states, respectively. 
The global electro-osmotic coefficient $L_{12}$ 
becomes negative under Condition (C). 
These conditions are marked in Fig.~\ref{fig5}(a). 
In all cases, the gradient of the flow field is localized 
to the vicinity of the surface; that is, the flow macroscopically 
behaves as a plug flow. 
While curve (A) varies almost monotonically with $y$, 
curve (B) is nonmonotonic, and curve  (C) is more complex. 
Under condition (C), 
the flow direction is positive near the surface, 
but changes at some distance from the wall, 
saturating at a negative value. The saturation value gives the
macroscopic electro-osmotic 
coefficient from Eq.~(\ref{eq:L12}). 
By contrast, curve (B) remains positive across the range. 
If the viscosity is homogenous and independent of the polymer concentration, 
the flow field is easily calculated from Eq.~(\ref{eq:lambda12}) as 
\begin{equation}
\lambda_{12}(y)
=\frac{\varepsilon}{4\pi\eta}\left\{\psi(y)-\psi_\mathrm{S}\right\}. 
\end{equation}
The overcharging potential  is necessary the nonmonotonic 
variations in (B) and (C). 
However, because $\psi_{\rm S}<\psi(y)$ everywhere, 
the overcharging potential alone cannot explain the negative $L_{12}$ 
given that $\eta$ is constant. 

If the electrostatic potential monotonically changes with $y$ 
as in condition (A), 
$\psi=\psi(y)$ is uniquely expressed by its 
inverse function $y=y(\psi)$. 
Then, 
Eq.~(\ref{eq:lambda12}) is given by 
\begin{equation}
\lambda_{12}(L)
=\frac{\varepsilon}{4\pi}\int^0_{\psi_\mathrm{S}}\eta(\psi')^{-1}d\psi',
\label{eq:lambda12_2}
\end{equation}
where $\eta(\psi)=\eta(\phi(y(\psi)))$ is also a unique function of 
$\psi$. 
The curves of $\eta(\psi)$ are plotted in Fig.~\ref{fig8}(b). 
Since $\eta(\psi)$ is positive, 
$\lambda_{12}(L)$ is also positive, 
indicating that the flow toward $E$ is maintained. 

When the overcharging potential arises, as in conditions (B) and (C), 
$y$ is a multivalued function of $\psi$, which invalidates 
Eq.~(\ref{eq:lambda12_2}). 
In this case, Eq.~(\ref{eq:lambda12}) becomes 
\begin{eqnarray}
\frac{4\pi}{\varepsilon}\lambda_{12}(L)&=&
\int_0^{y_\psi}
\frac{dy'}{\eta(y')}
\left.\frac{\partial\psi}{\partial y}\right|_{y'}
+\int_{y_{\psi}}^L
\frac{dy'}{\eta(y')}
\left.\frac{\partial\psi}{\partial y}\right|_{y'}
\nonumber\\
&=&
\int^{\psi_\mathrm{M}}_{\psi_\mathrm{S}|y<y_\psi}
\frac{d\psi'}{\eta(\psi')}
-\int^{\psi_\mathrm{M}}_{0|y>y_\psi}\frac{d\psi'}{\eta(\psi')}.
\label{eq:lambda12_3}
\end{eqnarray}
Here we should note that the paths of the two integrals in 
Eq.~(\ref{eq:lambda12_3}) differ from each other. 

According to linear analysis, 
the electrostatic potential profile 
may have multiple peaks.\cite{Chatellier_Joanny_1996}
The intensities of the peaks decay with increasing distance from the wall. 
We assume that the highest peak (nearest the wall) plays a dominant role 
in the electrokinetic flow 
and ignore the contributions of 
the remaining peaks.

\begin{figure}
\includegraphics[width=0.5\textwidth]{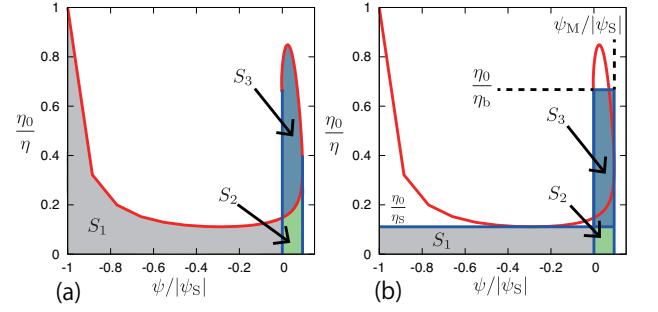}
\caption{
(a) Schematic for calculating $L_{12}$ 
in adsorption states from Eq.~(\ref{eq:lambda12_3}). 
$\psi$ and $1/\eta$ are parameterized with respect to $y$. 
(b) Approximate representation 
of $\psi$-$\eta^{-1}$ in (a). 
This approximation gives a simple form of $L_{12}$, Eq.~(\ref{eq:L12_est}). 
}
\label{fig9}
\end{figure}

Figure~\ref{fig9} is a schematic of
Eq.~(\ref{eq:lambda12_3}). 
When the electrostatic potential overcharges, 
the curve of $1/\eta (\psi)$ is divisible into three segments. 
These segments delineate three realms, with areas 
denoted by $S_1$, $S_2$, and $S_3$. 
Within the slit, the realms correspond to the ranges 
$S_1$: $0<y<y_0$, $S_2$: $y_0<y<y_\psi$, and $S_3$: $y_\psi<y<L$ 
(see Fig.~\ref{fig6}(b)). 
The first and second terms in Eq.~(\ref{eq:lambda12_3}) 
are given by $S_1+S_2$ 
and $S_2+S_3$, respectively. 
In terms of these areas, the electro-osmotic coefficient is given by 
$L_{12}=(S_1+S_2)-(S_2+S_3)=S_1-S_3$. 
If $S_1<S_3$, the macroscopic flow is inverted.

Using Eq.~(\ref{eq:lambda12_3}), we devise a simple method for estimating the 
electro-osmotic coefficient in 
adsorption states. 
The viscosity is assumed constant within each realm. 
More precisely, we assume that polymer 
concentration is fixed at $\phi=\phi_{\rm M}$ within the range 
$0<y<y_{\rm \psi}$ and at $\phi=\phi_{\rm b}$ in 
$y_{\rm \psi}<y<L$. 
These approximations are schematically represented 
in Fig.~\ref{fig9}(b). 
$S_1$ and $S_3$ are then approximated as 
\begin{eqnarray}
S_1&\approx& \frac{\varepsilon}{4\pi}\frac{-\psi_{\rm S}}{\eta_{\rm S}},\\
S_3&\approx& \frac{\varepsilon}{4\pi}
\left(\frac{1}{\eta_{\rm b}}-\frac{1}{\eta_
{\rm S}}\right)\psi_{\rm M}, 
\end{eqnarray}
where $\eta_{\rm S}=\eta_0(1+h\phi_{\rm M}/\phi_{\rm b})$ and 
$\eta_{\rm b}=\eta_0(1+h)$. 
Finally, we obtain 
\begin{eqnarray}
L_{12}\approx \frac{\eta_{\rm b}}{\eta_{\rm S}}
L_{12}^{\rm b}+\left(1-\frac{\eta_{\rm b}}{\eta_{\rm S}}\right)
L_{12}^{\rm M},
\label{eq:L12_est}
\end{eqnarray}
where 
$L_{12}^{\rm M}=-\varepsilon \psi_{\rm M}/(4\pi\eta_{\rm b})$ 
is the electro-osmotic coefficient estimated by 
the overcharging potential. 
The $L_{12}=0$ curve estimated by Eq.~(\ref{eq:L12_est}) 
is drawn in Fig.~\ref{fig7}. 
This curve is qualitatively consistent with the numerical solutions. 
In this estimation, the overcharging potential does not directly cause 
the inversion of electro-osmotic flow; 
formation of the highly viscous layer is also important.

\section{Summary and Remarks}

Applying a continuum model, we study electro-osmosis in polymer solutions. 
From numerical calculations and theoretical estimations, 
we elucidated the behaviors of the electro-osmosis 
in polymer solutions.
The dependence of viscosity on the polymer concentration  
plays an important role in our model. Our main results are summarized below.

\begin{itemize}
\item[(i)] 
Even if the polymer solution sandwiched between 
chemically repulsive walls is electrically neutral, 
electro-osmosis depends on the salt concentration. 
Decreasing the salinity suppresses the electro-osmosis. 
 
\item[(ii)]
In polyelectrolyte solutions, 
the formed adsorption layer effectively enlarges the 
viscosity in the vicinity of the surfaces. 
Consequently, 
electro-osmosis is suppressed much more strongly in polyelectrolyte  
than in neutral polymer solutions. 
If a sufficiently high proportion of the monomers are charged
and if the salt concentration is sufficiently low, 
the electro-osmotic flow can be inverted.

\item[(iii)]
We propose a simple equation 
for estimating 
the electro-osmotic coefficient in 
adsorption states (Eq.~(\ref{eq:L12_est})). 
This equation captures the essential features of the inversion 
of the electro-osmotic coefficient, shown in Fig.~\ref{fig7}. 
According to this expression, inversion is caused by two factors; 
enhancement of the viscosity by the near-surface adsorption layer 
and overshoot of the electrostatic potential.

\end{itemize}
We conclude this paper with the following remarks.

\begin{itemize}
\item[(1)]
Charge inversion and mobility reversal induced by multivalent 
electrolytes has been frequently 
reported.\cite{Grosberg_Nguyen_Shklovskii_2002} 
Grosberg {\it et al.}\cite{Grosberg_Nguyen_Shklovskii_2002} 
concluded that such phenomena depend on fluctuation correlations 
among the multivalent ions, 
which are excluded in usual Poisson-Boltzmann approaches. 
Our mean-field approach predicts that 
similar inversion phenomena occur in polyelectrolyte solutions. 
According to a molecular dynamics simulation, 
the phenomena occurs even in monovalent ions solutions 
confined within nanochannels.\cite{Qiao_Aluru_2004} 
The flow profiles obtained in the earlier study are quite similar to ours; 
near the surface, the flow is directed toward the electric field, 
but in the bulk, it is against the field.

\item[(2)]
This article 
considers only limited situations; 
The surfaces are assumed to chemically and electrostatically 
repel the polymers.
If the surfaces are chemically attractive, 
the adsorption is much enhanced by chemical
forces.\cite{Joanny_1999,Shafir_Andelman_2004}  
The electro-osmotic properties of these surfaces are equally interesting 
and important. 

\item[(3)] 
From the Onsager reciprocal relations, the electro-osmotic 
coefficient 
$L_{12}$ should equal $L_{21}$ in Eq.~(\ref{eq:Onsager2}). 
The latter represents the electric current due to the mechanical 
pressure difference. 
Interestingly, the Onsager coefficient $L_{21}$ is 
inverted when $f$ is large and $c_{\rm b}^+$ is sufficiently small.

\item[(4)] 
In the above numerical and theoretical analyses, 
the viscosity parameter $h$ is assumed constant, 
although in practice it depends on the 
fraction of charged monomers $f$ and the 
salt concentration $c_{\rm b}^+$.  
When $f$ is large and $c_{\rm b}^+$ is small, 
the solution viscosity increases 
(see Appendix \ref{app:B}). 
Our studies indicate that large $f$ and small $c_{\rm b}$ 
favor flow inversion. 
The same trends were observed for large $h$. 
If we set $h$ as a function of $f$ and $c_{\rm b}$, 
more dramatic changes would appear in the curves of $L_{12}$ 
against $f$ and $c_{\rm b}^+$. 
Although the $L_{12}$ and the phase diagrams 
would quantitatively alter, the qualitative trends, {\it i.e.},
suppression of  the electro-osmotic flow 
and inversion 
at large $f$ and a small $c_{\rm b}^+$, should remain intact.

\end{itemize}

\begin{acknowledgements}
YU is grateful to J.F. Joanny for helpful discussions since 
the start of this investigation, and
to D. Andelman for his valuable comments. 
YU is supported by a Grand-in-Aid for JSPS fellowship. 
This work was financially supported by the JSPS Core-to-Core Program 
``Non-equilibrium dynamics of soft matter and information" 
and KAKENHI (Nos. 23244088, 24540433, 25000002). 
The computational work was carried out using the facilities at the 
Supercomputer Center, Institute for Solid State Physics, University of Tokyo.

\end{acknowledgements}

\appendix

\section{Local equilibrium conditions for the components}
\label{app:A}

Because we apply an external field $E$ along the $x$ direction 
(see Fig.~\ref{fig1}), 
the total electrostatic potential is not $\psi(y)$ 
in Eq.~(\ref{eq:F_ele}), but instead is 
$\Psi(x,y)=\psi(y)-Ex$. 
Assuming the local equilibrium condition, the chemical potential 
of the $i$-th species is given by 
\begin{eqnarray}
\mu^i=k_{\rm B}T\ln (c^ia^3)+ez_i\Psi, 
\end{eqnarray}
where $z_i$ is the charge of the $i$-th ion. 
In the geometry of the investigated system, 
the diffusion flux of the ion, given by $\mbox{\boldmath $j$}^i
=-D^ic^i\nabla \mu^i$, is divided into two components: 
\begin{eqnarray}
\mbox{\boldmath $j$}^i
&=&j^i_y \mbox{\boldmath $e$}_y,  
+j^i_x \mbox{\boldmath $e$}_x, \\
j^i_y&=&-D^i c^+\frac{\partial}{\partial y}\left[k_{\rm B}T\ln (c^ia^3)
+ez_i\psi(y)\right]\\
j^i_x&=&D^i z_ic^iE.
\end{eqnarray}
Here $D^i$ is the diffusion constant of the $i$-th ion, and 
$\mbox{\boldmath $e$}_x$ and $\mbox{\boldmath $e$}_x$ 
are the unit vectors along the $x$ and $y$ axes, respectively. 
Because the system is confined by the walls at $y=0$ and $2L$, 
the diffusion flux along the $y$ direction vanishes 
at steady state. 
Thus, we obtain the Boltzmann distribution along the $y$ axis as 
$c^ia^3 \propto \exp\{-z_ie\psi/k_{\rm B}T\}$. 
On the other hand, the diffusion flux remains along the $x$ axis. 
Because the applied electric field is sufficiently weak and 
orthogonal to $-\nabla \psi$, 
it influences neither the concentration fields nor the polymer 
conformation.

\section{Scaling behaviors in polyelectrolyte solutions}
\label{app:B}

The scaling behaviors of polyelectrolyte solutions 
are known to widely differ from those of uncharged polymer solutions. 
At the overlap concentration $c^*$ in a polyelectrolyte solution,  
the monomer density 
inside the coil equals the overall monomer density in the 
solution.\cite{Dobrynin_Colby_Rubinstein_1995} 
In our notation, the overlap concentration in a theta solvent is determined by 
$c^*(1+2c_{\rm b}^+/c^*f)^{-3/2}\approx N^{-2}a^{-2}\ell_{\rm B}^{-1}f^{-2}$. 
 
In the low-salt or salt-free regime, 
the overlap concentration becomes
$c^*\approx (a^2\ell_{\rm B}Nf)^{-1}$. 
Conversely, it approaches 
$c^*\cong \{8(c_{\rm b}^{+})^3a^{-4}\ell_{\rm B}^{-2}f^{-7}N^{-4}\}^{1/5}$ 
in the high-salt regime. 
Between these two extremes, 
the overlap concentration decreases as $f$ increases. 
Given the same polymer length $N$, polyelectrolyte 
chains expand more than their uncharged 
counterparts. 

The viscosity of polyelectrolyte 
solutions also obeys scaling behaviors, which depend on the 
solvent quality and the polymer concentration regime. 
For example, 
the viscosity of a semidilute solution in a theta solvent 
is given by 
$\eta\approx\eta_0 N a \ell_\mathrm{B}^{1/2}fc^{1/2}
(1+2c_\mathrm{b}^+/fc)^{-3/4}$. 
If the salt is not dissolved or is insufficiently dilute, 
this expression approaches 
$\eta\approx \eta_0 N a\ell_\mathrm{B}^{1/2}fc^{1/2}$; that is, 
the viscosity is proportional to $c^{1/2}$ (Fuoss law). 
On the other hand, in highly saline conditions
the viscosity behaves as 
$\eta\approx \eta_0 N a\ell_\mathrm{B}^{1/2}
(c_\mathrm{b}^+)^{-3/4}f^{7/4}c^{5/4}$. 
The viscosity depends on the polymer concentration 
as $c^{5/4}$, identical to that of an uncharged polymer 
solution in a theta solvent, namely
$\eta\approx\eta_0 N(ca^3)^{1/(3\nu-1)}$ with $\nu=3/5$. 
Physically, this result implies
that electrostatic interactions in a polyelectrolyte solution 
are well screened by the salt.


\begin{thebibliography}{99}
\bibitem{Rubinstein_Papoian_2012}
M. Rubinstein and G. A. Papoian, Soft Matter {\bf 8}, 9265 (2012); 
and references therein. 
\bibitem{Oosawa_1971}
F. Oosawa, {\it Polyelectrolytes}, Marcel Dekker, Inc. (1971).
\bibitem{deGennes_1979}
P. G. de Gennes, {\it Scaling Concepts in Polymer Physics}, 
Cornell University Press (1979).
\bibitem{Joanny_Leibler_1990}
J.F. Joanny and L. Leibler, J. Phys. (France) {\bf 51}, 545 (1990).
\bibitem{Barrat_Joanny_1996}
J.L. Barrat and J.F. Joanny, Advances in Chemical Physics XCIV, 1 (1996).
\bibitem{Dobrynin_Rubinstein_2005}
A. V. Dobrynin and M. Rubinstein, Prog. Polym. Sci. {\bf 30}, 1049 (2005).
\bibitem{Igarashi_1991}
T. Igarashi {\it et al.}, Kobunshi Ronbunshu {\bf 48}, 751 (1991).
\bibitem{Liu_Yang_2009}
M. Liu and J. Yang, Microvascular Res. {\bf 78}, 14 (2009).
\bibitem{O'Brien_White_1978}
R. W. O'Brien and L. R. White, J. Chem. Soc. Faraday Trans. II {\bf 74}, 
1607 (1978).
\bibitem{Burgreen_Nakache_1964}
D. Burgreen and F. R. Nakache, J. Phys. Chem. {\bf 68}, 1084 (1964).
\bibitem{Rice_Whitehead_1965}
C. L. Rice and R. Whitehead, J. Phys. Chem. {\bf 69}, 4017 (1965).
\bibitem{Levine_Marriott_Robinson_1974}
B. S. Levine, J. R. Marriott, K. Robinson, 
J. Chem. Soc. Faraday Trans. II {\bf 71}, 1 (1974).
\bibitem{Joanny_Leibler_deGennes_1979}
J.F. Joanny, L. Leibler, P. G. de Gennes, 
J. Polym. Sci. Poym. Phys. Ed. {\bf 17}, 1073 (1979).
\bibitem{deGennes_1981}
P. G. de Gennes, Macromolecules {\bf 14}, 1637 (1981).
\bibitem{deGennes_1982}
P. G. de Gennes, Macromolecules {\bf 15}, 492 (1982).
\bibitem{Borukhov_Andelman_Orland_1995}
I. Borukhov, D. Andelman, H. Orland, Eurphys. Lett. {\bf 32}, 499 (1995).
\bibitem{Chatellier_Joanny_1996}
X. Ch\^atellier and J.F. Joanny, J. Phys. II France {\bf 6} 1669 (1996).
\bibitem{Borukhov_Andelman_Orland_1998}
I. Borukhov, D. Andelman, H. Orland, Macromolecules {\bf 31}, 1665 (1998).
\bibitem{Joanny_1999}
J.F. Joanny, Eur. Phys. J. B {\bf 9}, 117 (1999).
\bibitem{Shafir_Andelman_Netz_2003}
A. Shafir, D. Andelman, R. R. Netz, J. Chem. Phys. {\bf 119}, 2355 (2003).
\bibitem{Shafir_Andelman_2004} 
A. Shafir and D. Andelman, Phys. Rev. E {\bf 70}, 061804 (2004).
\bibitem{Vink_1988}
H. Vink, J. Chem. Soc., Faraday Trans. I {\bf 84}, 133 (1988).
\bibitem{Bello_1994}
M. S. Bello {\it et al.}, Electrophoresis {\bf 15}, 623 (1994).
\bibitem{Otevrel_Kleparnik_2002}
M. Otev\v rel and K. Klep\'arn\'ik, Electrophoresis {\bf 23}, 3574 (2002).
\bibitem{Zhao_Zholkovskij_Masliyah_Yang_2008}
C. Zhao, E. Zholkovskij, J. H. Masliyah, C. Yang, 
J. Colloid Interface Sci. {\bf 326}, 503 (2008).
\bibitem{Olivares_VeraCandioti_Berli_2009}
M. L. Olivares, L. Vera-Candioti, C. L. A. Berli, 
Electrophoresis {\bf 30}, 921 (2009).
\bibitem{Zhao_Yang_2011}
C. Zhao and C. Yang, J. Non-Newtonian Fluid Mech. {\bf 166}, 1076 (2011).
\bibitem{Dobrynin_Colby_Rubinstein_1995}
A. V. Dobrynin, R. H. Colby, M. Rubinstein, 
Macromolecules {\bf 28}, 1859 (1995). 
\bibitem{Grosberg_Nguyen_Shklovskii_2002}
A. Yu. Grosberg, T. T. Nguyen, B. I. Shklovskii, 
Rev. Mod. Phys. {\bf 74}, 329 (2002).
\bibitem{Qiao_Aluru_2004}
R. Qiao, N. R. Aluru, Phys. Rev. Lett. {\bf 92}, 198301 (2004).
\end{thebibliography}
\end{document}